# Direct search for a ferromagnetic phase in a heavily overdoped nonsuperconducting copper oxide

J. E. Sonier[a,b,1], C. V. Kaiser[a], V. Pacradouni[a], S. A. Sabok-Sayr[a], C. Cochrane[a], D. E. MacLaughlin[c], S. Komiya[d], and N. E. Hussey[e]

[a]Department of Physics, Simon Fraser University, Burnaby, British Columbia V5A 1S6, Canada

[b]The Canadian Institute for Advanced Research, Toronto, Ontario M5G 1Z8, Canada

[c]Department of Physics and Astronomy, University of California, Riverside, CA 92521, USA

[d]Central Research Institute of Electric Power Industry, Yokosuka, Kanagawa 240-0196, Japan

[e]H. H. Wills Physics Laboratory, University of Bristol, Bristol BS8 1TL, UK

**The doping of charge carriers into the $CuO_2$ planes of copper oxide Mott insulators causes a gradual destruction of antiferromagnetism and the emergence of high-temperature superconductivity. Optimal superconductivity is achieved at a doping concentration *p* beyond which further increases in doping cause a weakening and eventual disappearance of superconductivity. A potential explanation for this demise is that ferromagnetic fluctuations compete with superconductivity in the overdoped regime. In this case a ferromagnetic phase at very low temperatures is predicted to exist beyond the doping concentration at which superconductivity disappears. Here we report on a direct examination of this scenario in overdoped $La_{2-x}Sr_xCuO_4$ using the technique of muon spin relaxation. We detect the onset of static magnetic moments of electronic origin at low temperature in the heavily overdoped nonsuperconducting region. However, the magnetism does not exist in a commensurate long-range ordered state. Instead it appears as a dilute concentration of static magnetic moments. This finding places severe restrictions on the form of ferromagnetism that may exist in the overdoped regime. Although an extrinsic impurity cannot be absolutely ruled out as the source of the magnetism that does occur, the results presented here lend support to electronic band calculations that predict the occurrence of weak localized ferromagnetism at high doping.**

Attempts within the framework of standard theories have failed to explain how high-temperature superconductivity emerges from charge carrier doping of an antiferromagnetic (AF) Mott insulator (1). The conventional Bardeen-Cooper-Schrieffer (BCS) theory of low-temperature superconductors (2) assumes that above the critical transition temperature ($T_c$) the electrons form a Landau Fermi liquid, and that superconductivity arises from pair condensation of the associated low-energy excitations (quasiparticles). What is clear in the case of copper oxides is that over the initial doping range where superconductivity first appears, ordinary Landau Fermi liquid theory does not apply. This has prompted theories in which the properties of the ordinary Fermi liquid are hidden (3), or alternatively only some properties of the Fermi liquid persist (4).

A BCS-type theory may be applicable in the heavily overdoped region, where an ordinary Fermi liquid is observed (5-7). However, the recent proposal by Kopp *et al.* (8) that ferromagnetic (FM) fluctuations compete with *d*-wave superconductivity is a challenge to the notion that overdoped copper oxides strictly conform to such conventional wisdom. The primary motivation for their hypothesis is a strong upturn in the magnetic susceptibility immediately above $T_c$ for doping levels greater than $p \sim 0.19$ (9-14). A tendency toward FM order for high charge doping (Fig. 1) is supported by electronic band calculations for supercells of $La_{2-x}Ba_xCuO_4$ (15). However, these calculations favour the appearance of weak ferromagnetism about concentrated regions of the Ba-dopant atom, rather than the emergence of long-range FM order.

The muon spin relaxation (μSR) method is similar to nuclear magnetic resonance (NMR), but is a more sensitive probe of static or slowly fluctuating magnetism. Furthermore, μSR exploits muon beams possessing a naturally created ~ 100 % spin polarization, and hence unlike NMR does not require constant or time-varying external magnetic fields. Zero-field (ZF) μSR experiments on $La_{2-x}Sr_xCuO_4$ (LSCO) above 2 K show the absence of static electronic moments for $p > 0.12$ (16, 17), which we have confirmed by measurements on LSCO single crystals with Sr content $x$ = 0.15, 0.166, 0.176, 0.196, 0.216, 0.24 and 0.33, with corresponding doping concentrations $p$ = 0.15, 0.166, …0.33 (henceforth referred to as LSCO15, LSCO166, …LSCO33). However, the onset of static FM order is expected to occur at a very low temperature, determined by the weak interlayer coupling of the $CuO_2$ planes (8). Consequently, we have extended the ZF-μSR measurements down to 0.02 K using a dilution refrigerator.

**Results and Discussion**

The time evolution of the muon spin polarization $P(t)$ is the physical quantity that is directly measured in the μSR experiments. Relaxation of $P(t)$ is caused by a distribution of static and/or fluctuating local magnetic fields. Representative ZF-μSR spectra for a single LSCO24 crystal are shown in Fig. 2*a* for the initial polarization $P(0)$ directed parallel to the crystallographic *c*-axis. There is a nonrelaxing contribution from muons stopped in the pure silver sample holder that accounts for about 70 % of the signal. The remaining signal is temperature independent, and hence is dominated by the distribution of dipolar fields from the nuclear magnetic moments.

The ZF-μSR measurements on LSCO33 were performed simultaneously on two smaller single crystals. In contrast to LSCO24, the ZF-μSR spectrum for LSCO33 shows an increased relaxation at temperatures below $T \sim 0.9$ K (Fig. 2*b*), indicating the occurrence of magnetic moments of electronic origin. To determine whether these moments are static or dynamic, a longitudinal field (LF) $B_L$ was applied parallel to $P(0)$ to decouple the muon spin from the static internal fields. Fast field fluctuations in a Gaussian distribution result in a LF exponential relaxation rate given by (18)

$$\lambda(B_L) = \frac{\Lambda_{ZF}}{1+(\gamma_\mu B_L \tau)^2} , \qquad [1]$$

where $1/\tau$ is the fluctuation rate. Alternatively, fast field fluctuations in a Lorentzian distribution result in a LF exponential relaxation rate given by (18)

$$\lambda(B_L) = \frac{\Lambda_{ZF}}{\sqrt{1+(\gamma_\mu B_L \tau)^2}} . \qquad [2]$$

Fig. 2*c* shows LF-μSR spectra of LSCO33 at a field of $B_L$ = 14 G. The LF-μSR spectrum at 3.2 K is well described by a static Gaussian LF Kubo-Toyabe function, which indicates that the relaxation observed at zero field is caused by the static internal field distribution of the nuclear dipoles (which are dense and randomly oriented). The LF-μSR spectrum at 0.02 K is described by the product of the same static Gaussian LF Kubo-Toyabe function and an exponential relaxation function with a relaxation rate of $\lambda(B_L) = 7.44 \times 10^{-4}$ μs$^{-1}$. In Fig. 3 we compare this to the value of $\lambda(B_L = 14$ G$)$ calculated from Eq. **1** and Eq. **2**, where $\Lambda_{ZF}$ = 0.12(1) μs$^{-1}$ at 0.02 K. This plot puts an upper limit on the fluctuation rate of $1/\tau \leq 9.5 \times 10^4$ Hz. Hence the anomalous electronic moments detected at 0.02 K are frozen in time.

The static magnetism in LSCO33 at zero field cannot be lingering AF spin correlations from the parent insulator. While neutron scattering measurements (19) show that AF correlations vanish below $x = 0.30$, beyond $x \sim 0.12$ these are known (16) to fluctuate on a much shorter time scale than the μSR time window ($10^{-12}$ to $10^{-4}$ s). Here we point out that minority phases of lower Sr concentration in overdoped samples usually exceed $x = 0.15$, as this is the most stable phase of superconducting LSCO single crystals. It is also apparent that the enhanced relaxation rate below 0.9 K does not signify a phase of long-range FM order, which would manifest itself as an oscillating ZF-μSR signal with a frequency proportional to the average field at the muon site. Instead the ZF-μSR spectra for LSCO33 are reasonably described by the polarization function (Fig. 2b)

$$P(t) = 0.35 \exp[-\Lambda_{ZF}(T)\, t] \cdot G_{nuclear}(t) + 0.65. \qquad [3]$$

The first term describes the depolarization associated with muons implanted in the sample, and the second term is from muons stopping in the sample holder. The temperature-independent relaxation function $G_{nuclear}(t)$ is dominated by the nuclear moments. The additional exponential relaxation rate that appears below 0.9 K (Fig. 2d) indicates a dilute system of static electronic moments. At 0.02 K the value of $\Lambda_{ZF}(T)$ corresponds to a Lorentzian distribution of local fields with a half width at half maximum (HWHM) of 1.4 G. Of the minority phases of the raw materials, $La_2O_3$ and $SrCO_3$ are nonmagnetic, and CuO is antiferromagnetic, but with a Neel temperature greatly exceeding 0.9 K. Since an additional distinct relaxation rate is not observed below 0.9 K, all of the implanted muons apparently sense the static moments, indicating that the magnetism is present throughout both LSCO33 crystals. This and the absence of such magnetism in the LSCO24 single crystal makes it unlikely that it is caused by an extrinsic impurity.

To gain further insight into the origin of the magnetism we have also performed bulk magnetic susceptibility measurements in external magnetic fields up to $H = 7$ T. The bulk dc-magnetic susceptibility of LSCO has been extensively studied (9, 10, 12-14, 20, 21), and below $x \sim 0.19$ is given by (20)

$$\chi(x,T) = \chi_0(x) + \chi^{2D}(x,T), \qquad [4]$$

where $\chi_0(x)$ is the temperature-independent uniform susceptibility and $\chi^{2D}(x,T)$ is the effective susceptibility of the $Cu^{2+}$ spin sublattice. The functional form of $\chi^{2D}(x,T)$ is that of a two-dimensional (2D) Heisenberg antiferromagnet. Above $x \sim 0.19$ there is an anomalous occurrence of Curie paramagnetism, such that

$$\chi(x,T) = \chi_0(x) + \chi^{2D}(x,T) + \frac{C(x)}{T}, \qquad [5]$$

where the Curie constant $C(x)$ increases with increased doping (10, 13, 14), but subsequently decreases (10, 22) above $x \sim 0.26$ (*i.e.* beyond the superconducting phase). The onset of the Curie term is accompanied by a saturation of the superfluid density in the superconducting phase (23) and a monotonic increase in the width of the internal magnetic field distribution with increasing hole doping in an external field at temperatures above $T_c$ (24). These observations indicate that the excess $Sr^{2+}$ ions beyond $x \sim 0.19$ do not enhance the density of superconducting carriers, but instead induce a new form of magnetism. MacDougall *et al.* (25) have suggested that the overdoped $Sr^{2+}$ ions induce a local staggered magnetization, similar to that caused by controlled impurity substitutions on the Cu(2) sites (26). In this scenario the decrease of $C$ above $x \sim 0.26$ may be caused by screening of the $Sr^{2+}$ induced local moments by surrounding conduction electrons, similar to the Kondo-like effect that is believed to occur in $Li^+$-substituted $YBa_2Cu_3O_{6+y}$ at high doping (27). However, this explanation is contrary to the development of frozen moments at $x = 0.33$, which instead suggests the emergence of a new kind of magnetism.

In Fig. 4 we show the temperature dependence of $\chi$ for LSCO24 and LSCO33 at magnetic fields applied parallel to the $CuO_2$ planes. In striking contrast to LSCO24, the product $\chi T$ for LSCO33 exhibits a linear temperature dependence over a wide range of temperature (Fig. 4*b*). According to Eq. **5**, this indicates that $\chi^{2D}(x,T)$ and remnant AF correlations are absent in LSCO33. On the other hand, the susceptibility of LSCO33 develops an appreciable dependence on field below $T \sim 15$ K (inset of Fig. 4*a*), although the magnetization $M$ at 2 K does not reach saturation even at 7 T (Fig. 4*c*). This field dependence is consistent with spin freezing at lower temperature. A simultaneous fit of $\chi(T)$ and $M(H)$ between $T = 15$ and 150 K to a Brillouin function yields an effective moment of 2.86 $\mu_B$, and a density of 0.00312 moments per tetragonal unit cell of LSCO33. This corresponds to a Curie constant of $C = 3.1 \times 10^{-6}$ emu K/g. The magnetic field distribution sensed by the muon ensemble for static moments of this kind is approximately a Lorentzian function with a HWHM of $\sim 1.6$ G (Fig. S1), irrespective of whether the moments are randomly oriented or parallel to the $c$-axis. As this is close to the value of 1.4 G deduced from $\Lambda_{ZF}$ at 0.02 K (Fig. 2*d*), the electronic moments responsible for the Curie paramagnetism are the same moments that freeze below 0.9 K.

The in-plane electrical resistivity of LSCO33 above 60 K can be described by a non-Fermi liquid temperature dependence of the form $\rho_{ab}(T) = \rho_{ab}(0) + AT^{5/3}$ (Fig. 4*d*). The $T^{5/3}$ power law and the Curie behaviour of the magnetic susceptibility well above the spin freezing temperature of 0.9 K, are indicative of weak itinerant-electron ferromagnetism (29). Below 60 K the temperature dependence of $\rho_{ab}(T)$ becomes stronger, and the conventional $T^2$ Fermi-liquid behaviour is observed over the range 4.5 K to 50 K (inset of Fig. 4*d*). The situation resembles the weak itinerant superconducting ferromagnet $Y_4Co_3$, which exhibits $T^2$ resistivity extending above the Curie temperature and a crossover to $T^{5/3}$-behaviour at higher temperatures (30). Thus, our findings lend support to electronic band calculations (15) and a scenario whereby substitution of $Sr^{2+}$ for $La^{3+}$ at high doping does not enhance superconductivity, but instead induces weak itinerant ferromagnetism "localized" primarily on nearby Cu atoms. The creation of local moments by $Sr^{2+}$ ions in excess of $x \sim 0.19$, in the form of regions of staggered magnetization and/or weak ferromagnetism, is consistent with the saturation of the superfluid density (23) and the decline of $T_c$ in the heavily overdoped region. While the present study establishes the absence of a long-range FM phase beyond the SC "dome", a systematic study as a function of doping is needed to definitively rule out an extrinsic impurity as the source of the observed magnetism at $x = 0.33$. This currently awaits the availability of additional sizeable single crystals in the heavily overdoped nonsuperconducting regime.

**Materials and Methods**
The $La_{2-x}Sr_xCuO_4$ (LSCO) single crystals were grown by a travelling-solvent floating zone (TSFZ) technique (31). $La_2O_3$, $SrCO_3$, and CuO powders were mixed with the cation ratio of La : Sr : Cu = (2-$x$) : $x$ : 1.02 for each $x$ and calcined at 780 °C for 12 hours, and then reground and calcined at 920 °C for 12 hours 3 times. The 2 % excess Cu works to compensate the loss of Cu by evaporation during the TSFZ process. The raw powder was pressed into a rod shape, and sintered at 1200 °C for 15 hours. For the solvent material, the above raw powder and additional CuO were mixed with a cation ratio of (La, Sr) : Cu = 2 : 3. The TSFZ growth was operated at a growth rate of less than 1 mm/h in dry flowing air. After growth, the rod was cut into platelets, and annealed at 800 to 900 °C in appropriate oxygen partial pressure to remove oxygen defects according to the oxygen nonstoichiometry data (32).

We note that there is some variation in the literature on the precise Sr content $x$ at which superconductivity ceases, because samples in the doping range $x = 0.27$ to $0.30$ often contain underdoped superconducting regions. The LSCO33 sample reported on here was annealed for two weeks under extreme oxygen pressure (400 atm) to homogenize the oxygen content and accordingly exhibits no (resistivity) trace of superconductivity down to 0.1 K.

**ACKNOWLEDGMENTS.** We thank S. Chakravarty, B. Barbiellini, T. Jarlborg, R. Cywinski, D. Leznoff and T. Adachi for informative discussions. We also thank M. Lees, R. Liang, N. Mangkorntang, M. Nohara, H. Takagi and the facility personnel of TRIUMF's Centre for Molecular and Materials Science for technical and/or experimental assistance. This work was supported by the Natural Sciences and Engineering Research Council of Canada, the Canadian Institute for Advanced Research, and the Engineering and Physical Sciences Research Council of the United Kingdom.

Author contributions: S.K. and N.H. provided the LSCO single crystals. N.H. was involved in the resistivity measurements. J.E.S., C.V.K., V.P., S.A.S. and C.C. performed the μSR experiments. C.V.K. carried out the magnetization measurements. All authors contributed to the making of this manuscript.

[1]To whom correspondence may be addressed. E-mail: jsonier@sfu.ca.

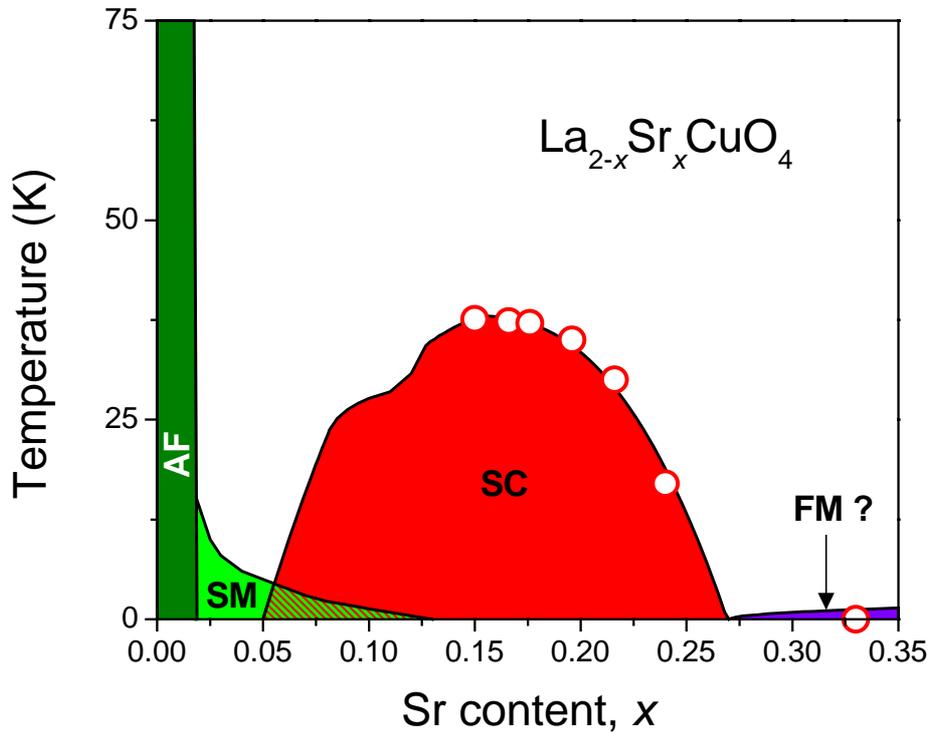

**Fig. 1.** Schematic phase diagram of $La_{2-x}Sr_xCuO_4$. The dark green area represents the region in which static long-range AF order is observed. In the light green region inhomogeneous static magnetism (SM) occurs that coexists with superconductvity between $0.05 < x < 0.13$. The red region represents the superconducting (SC) phase and the red open circles are the $T_c$ values of the single crystals studied here by μSR. The purple area beyond the SC "dome" represents the long-range static FM phase predicted by Kopp *et al.* (8). FM fluctuations associated with this phase are predicted to compete with superconductivity above $x \sim 0.19$. Alternatively, FM clusters (16) may appear above $x \sim 0.19$ and freeze into a dilute FM phase beyond $x = 0.27$, which is also represented by the purple area.

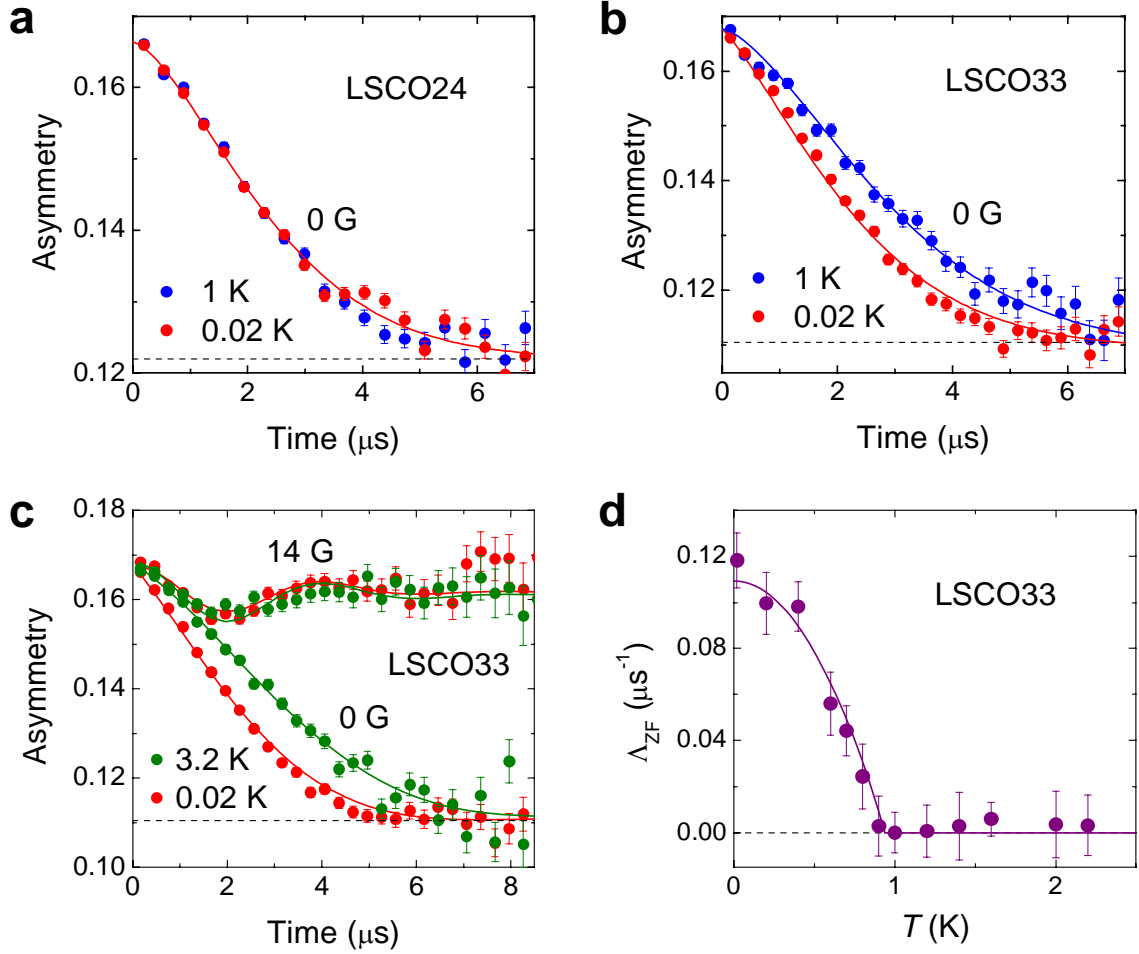

**Fig. 2.** Comparison of ZF-μSR spectra for heavily overdoped superconducting and nonsuperconducting $La_{2-x}Sr_xCuO_4$. (*a*) ZF-μSR asymmetry spectra for a plate-like LSCO24 single crystal of mass 127.5 mg recorded at temperatures 0.02 K and 1 K, with the initial muon spin polarization $P(0)$ parallel to the *c*-axis. The horizontal dashed line (also in *b*) indicates the time-independent contribution from muons stopping in the pure silver sample holder. The solid curve through the data is a fit to $a_0 P(t) = a_0 [0.27 G_{\text{nuclear}}(t) + 0.73]$, where $a_0$ is the initial amplitude and $G_{\text{nuclear}}(t) = \exp[-(\Delta t)^p]$, with $\Delta = 0.366$ μs$^{-1}$ and $p = 1.505$. (*b*) ZF-μSR spectra for a LSCO33 sample consisting of two single crystals. Crystal #1 is plate-like with a mass of 33.7 mg and crystal #2 is cylindrical-wedge shaped with a mass of 49.1 mg. The ZF-μSR spectra were recorded at temperatures 0.02 K and 1 K with $P(0)$ parallel to the $CuO_2$ planes of crystal #1, and at an arbitrary angle with respect to the *c*-axis of crystal #2. (*c*) ZF-μSR and LF-μSR spectra for LSCO33 at 0.02 K and 3.2 K. The LF-μSR spectra were recorded with a field of $B_L = 14$ G applied parallel to the direction of $P(0)$. The LF-μSR spectra are fit to the product of a common static Gaussian LF relaxation function (to account for the contribution of the nuclear dipoles) and an exponential relaxation function. The exponential relaxation rate is $\lambda(B_L) = 1.26 \times 10^{-7}$ μs$^{-1}$ and $\lambda(B_L) = 7.44 \times 10^{-4}$ μs$^{-1}$ at 3.2 K and 0.02 K, respectively. (*d*) Temperature dependence of the ZF relaxation rate $\Lambda_{ZF}$ from fits of the ZF-μSR asymmetry spectra for LSCO33 to $a_0 P(t) = a_0 \{0.35 \exp[-\Lambda_{ZF}(T)\, t] \cdot G_{nuclear}(t) + 0.65\}$, with $\Delta = 0.297$ μs$^{-1}$ and $p = 1.514$.

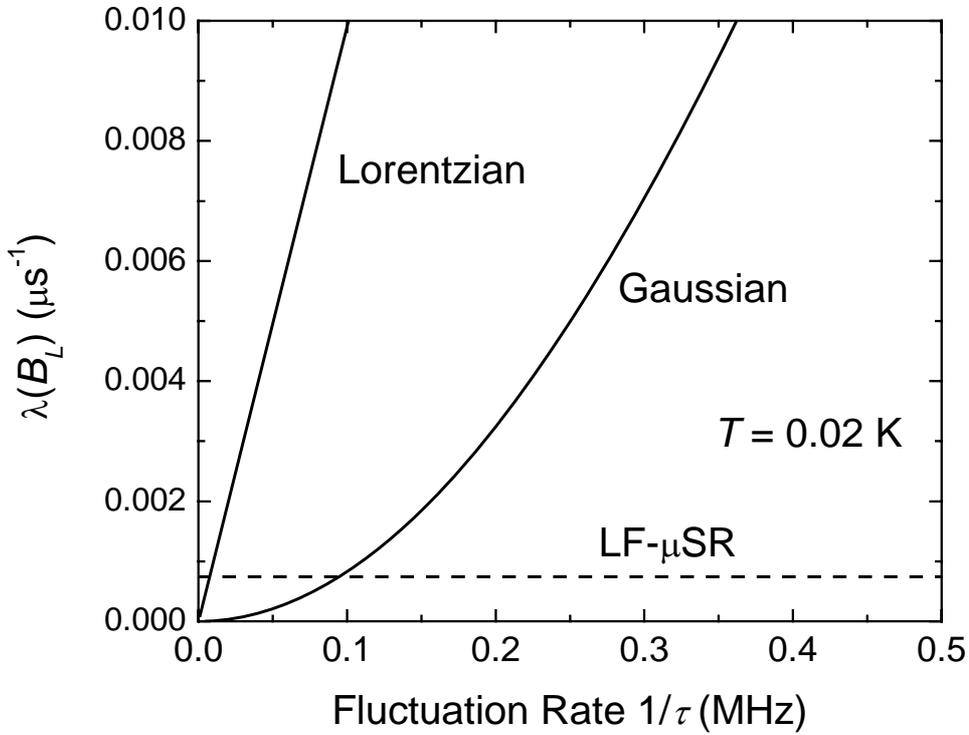

**Fig. 3.** Exponential decay of the muon spin polarization in a LF of $B_L = 14$ G. The solid curves are calculated from Eq. **1** and Eq. **2**, respectively, using the ZF relaxation rate of $\Lambda_{ZF} = 0.12(1)$ μs$^{-1}$ at 0.02 K. The dashed line indicates the relaxation rate of the exponential component of the LF-μSR spectrum at 0.02 K.

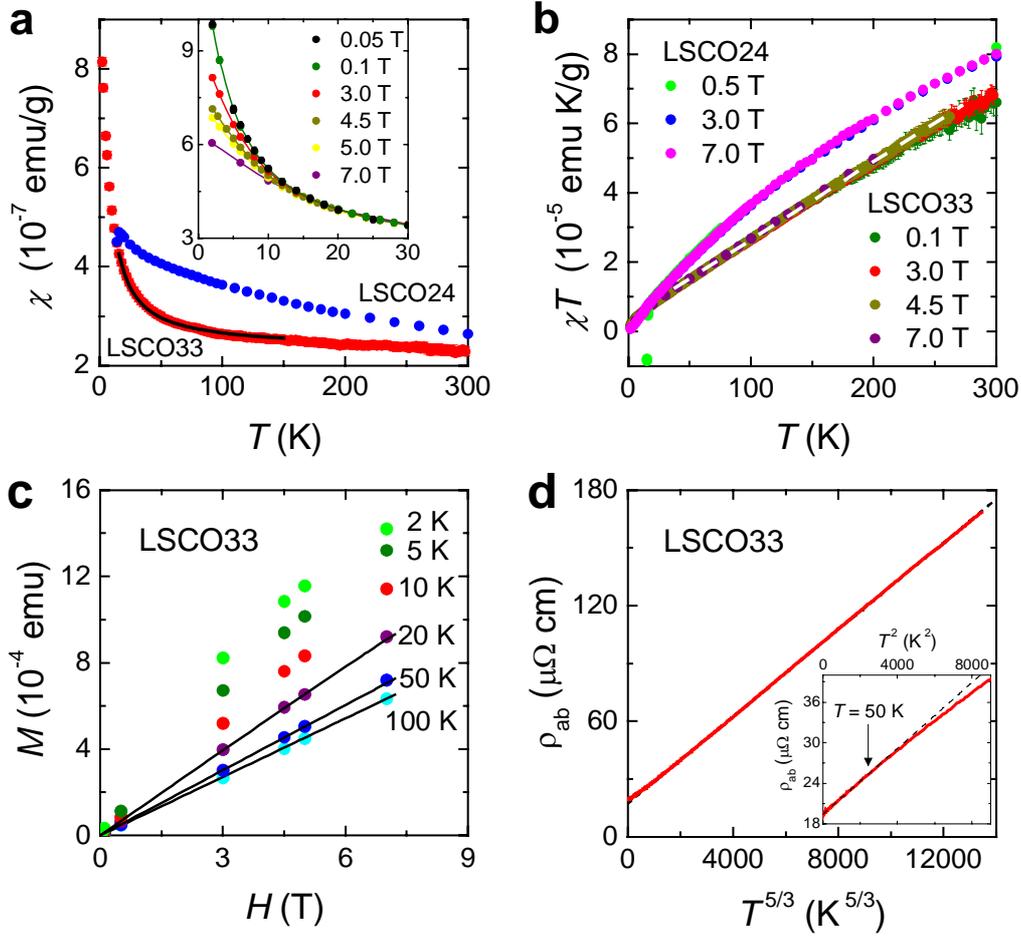

**Fig. 4.** Bulk dc-magnetic susceptibility and electrical resistivity. (*a*) Bulk magnetic susceptibility of the LSCO24 single crystal (blue circles) and LSCO33 single crystal #1 (red circles) for a field of 3 T applied parallel to the $CuO_2$ planes. The inset shows the susceptibility for fields $H$ = 0.05, 0.1, 3.0, 4.5, 5.0, and 7.0 T. (*b*) Temperature dependence of the product $\chi T$ at different applied magnetic fields. The white dashed straight line through the data of LSCO33 is a guide to the eye. (*c*) The magnetization of LSCO33 versus applied field $H$ at temperatures from $T$ = 2 to 100 K. The solid black curves are simultaneous fits of the data at $T$ = 20, 50 and 100 K to a Brillouin function, yielding $N$ = 1.82 × $10^{18}$ paramagnetic moments per gram, each with an effective moment of 2.86 $\mu_B$. The corresponding susceptibility at $H$ = 3 T between $T$ = 15 and 150 K is shown in the main panel of (*a*), as a solid black curve. (*d*) Zero-field in-plane electrical resistivity of LSCO33 over the temperature range 4.5 K to 300 K (28) plotted versus $T^{5/3}$. The black-dashed line that is barely visible indicates the deviation of the resistivity from $T^{5/3}$ below 60 K. The inset shows the in-plane resistivity data for LSCO33 below 95 K plotted as a function of $T^2$. The black-dashed line indicates the deviation of the resistivity from $T^2$ behaviour above 50 K.

# Supporting Information

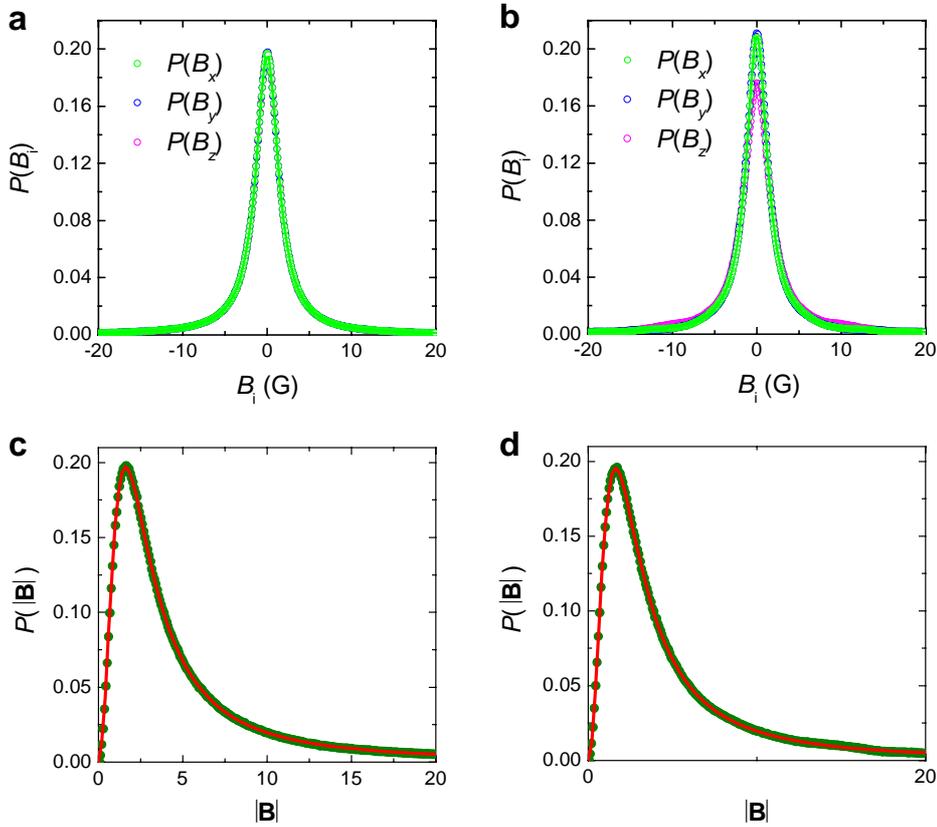

**Fig. S1** Simulations of the distribution of the spatial components and magnitude of the local magnetic field at the muon site for frozen magnetic moments in LSCO33. The field distributions are for 2.86 $\mu_B$ magnetic moments randomly distributed with 0.00312 moments per tetragonal unit cell of LSCO33. The muon site in the LSCO33 tetragonal unit cell is (0.253$a$, 0.0$b$, 0.162$c$), where $a = b = 3.78$ Å and $c = 13.2$ Å are the lattice constants. (*a*) shows the distribution of field components for randomly oriented moments. (*b*) shows the distribution of field components for an equal number of moments aligned parallel and anti-parallel to the *c*-axis. The curves in each figure show fits to a Lorentzian distribution of field components. (*c*) The distribution of the field magnitude for the case of randomly oriented moments. The red curve is a fit assuming an isotropic Lorentzian distribution of field components with a HWHM of 1.62 G. (*d*) The distribution of the field magnitude for the case of the moments parallel and anti-parallel to the *c*-axis. The red curve is a fit assuming an isotropic Lorentzian distribution of field components with a HWHM of 1.63 G. Note that aligning all of the moments parallel to the *c*-axis increases the HWHM of the Lorentzian distribution slightly to 1.66 G.